\documentclass{PoS}

\usepackage{amssymb}
\usepackage{amsmath}
\usepackage{natbib}
\bibpunct{(}{)}{;}{a}{}{,} 
\usepackage{graphicx}
\usepackage{color}
\usepackage{epstopdf}
\usepackage{wasysym}
\usepackage{booktabs}

\title{Bow shock sources close to the Galactic centre}

\ShortTitle{Bow shock sources}

\author{\speaker{Michal Zaja\v{c}ek}\\
        Center for Theoretical Physics, Polish Academy of Sciences, Al. Lotnikow 32/46, 02-668 Warsaw, Poland\\
        I. Physikalisches Institut der Universit\"at zu K\"oln, Z\"ulpicher Strasse 77, D-50937 K\"oln, Germany\\
        Max-Planck-Institut f\"ur Radioastronomie (MPIfR), Auf dem H\"ugel 69, D-53121 Bonn, Germany
        E-mail: \email{zajacek@ph1.uni-koeln.de}}

\author{Andreas Eckart, Seyedeh Elaheh Hosseini\\
        I. Physikalisches Institut der Universit\"at zu K\"oln, Z\"ulpicher Strasse 77, D-50937 K\"oln, Germany\\
        Max-Planck-Institut f\"ur Radioastronomie (MPIfR), Auf dem H\"ugel 69, D-53121 Bonn, Germany}

\abstract{We provide an up-to-date summary of the current observational and theoretical studies of stellar bow-shock sources close to the Galactic centre. The symmetry axis of a bow shock provides the information on the relative motion of the star with respect to the ambient medium, while the photometry and spectroscopy in NIR domain give information about the 3D motion of the star. Hence, it is possible from this data to obtain an estimate on the motion of the ambient medium. In combination with the estimate of the bow-shock size, it is possible to infer the valuable information on the density of the hot accretion flow close to the Galactic centre. In particular, we outline a statistical method to determine the ambient density slope based on either multiple bow-shock detections for one star along its orbit or multiple bow-shock detections for several sources at different distances from Sgr~A*.}

\FullConference{Accretion Processes in Cosmic Sources --- II - APCS2018\\
		3--8 September 2018\\
		Saint Petersburg, Russian Federation}

\begin{document}

\section{Introduction}

The Galactic centre environment represents a unique astrophysical laboratory where one can study the dynamical effects inside one of the densest stellar clusters on one hand \citep{2005PhR...419...65A,2013degn.book.....M} and it also allows us to study the mutual interaction between stars and the gaseous-dusty medium inside the sphere of influence of compact radio source Sgr~A* on the other hand \citep[see][for overviews]{2010RvMP...82.3121G,2017bhns.work..237Z,2017FoPh...47..553E}.

The Galactic centre is characterized by an overall increase in the NIR and MIR surface brightness along the Galactic plane \citep{1968ApJ...151..145B,1969ApJ...157L..31B,1969ApJ...157L..97L}. The NIR emission is dominated by the photosphere emission of individual stellar sources that constitute a crowded stellar field, while the MIR emission originates in both the diffuse gaseous-dusty circumnuclear medium and the dusty circumstellar envelopes of a few stellar sources.   

 In terms of stellar populations, there are both late-type stars whose number density forms a flat core towards the centre and early-type stars that form a steeper cusp \citep{2009A&A...499..483B}. There are clear signatures that recent \textit{in situ} star-formation event took place inside the sphere of influence of the supermassive black hole (SMBH) associated with Sgr~A*. This is mainly based on the three populations of massive young OB stars in the innermost parsec \citep{2014A&A...567A..21S}:
\begin{itemize}
  \item scattered population at the distance of $0.04$ up to $1.2$ parsecs (hereafter pc),
  \item disc-like population at $0.04\,{\rm pc}$ up to $0.5\,{\rm pc}$,
  \item a population of fast-moving, lighter B stars inside $0.04\,{\rm pc}$, which is also known as the S cluster \citep{1996Natur.383..415E,1997MNRAS.284..576E}.
\end{itemize}
These stars have large mass-loss rates of as much as $\dot{M}_{\rm w}\approx 10^{-7}\,M_{\odot}{\rm yr^{-1}}-10^{-5}\,M_{\odot}{\rm yr^{-1}}$ and high-velocity stellar winds $v_{\rm w}\approx 10^3\,{\rm km\,s^{-1}}$ and they can, therefore, strongly interact with the surrounding medium.

Inside the sphere of influence, $r_{\rm inf}\sim 1.7\,{\rm pc}$ for Sgr~A*, stars orbit the SMBH approximately with Keplerian velocities,

\begin{equation}
  v_{\star}\simeq 415\,\left(\frac{M_{\bullet}}{4\times 10^6\,M_{\odot}}\right)^{1/2} \left(\frac{r}{0.1\,{\rm pc}}\right)^{-1/2}\,{\rm km\,s^{-1}}\, \text{ for }\, r\lesssim r_{\rm inf},
  \label{eq_orb_vel}
\end{equation}
where $r$ is the distance from Sgr~A*. The basic condition for the bow-shock formation in the circumnuclear gaseous-dusty medium is that the relative velocity of the star with respect to the ambient medium $\mathbf{v}_{\rm rel}=\mathbf{v}_{\star}-\mathbf{v}_{\rm a}$ is larger than the local speed at which signals and perturbations propagate in the medium. For the non-magnetized or weakly magnetized medium, the wave speed is the local sound speed $c_{\rm s}(r)$,

\begin{equation}
  c_{\rm s}=482\,\left(\frac{\gamma}{1.4}\right)^{1/2}\left(\frac{T_{\rm a}}{10^7\,{\rm K}}\right)^{1/2}\left(\frac{\mu_{\rm HII}}{0.5}\right)^{-1/2}\,{\rm km\,s^{-1}},
  \label{eq_sound_speed}
\end{equation}
where we have assumed that the ambient medium is fully ionized. The temperature of the ambient medium $T_{\rm a}$ is based on the X-ray observations on the scale of the Bondi radius \citep{2003ApJ...591..891B,2013Sci...341..981W}. By dividing Equations~\ref{eq_orb_vel} and \ref{eq_sound_speed}, we essentially obtain the Mach speed $\mathcal{M}\equiv v_{\rm rel}/c_{\rm s}$, which is close to unity at the Bondi radius. An additional inflow or outflow, especially approximately perpendicular to the orbital velocity, can increase the relative velocity, which sets the Mach speed above unity and the bow-shock formation can occur in the ambient non-magnetized medium. It should be noted that even for the subsonic relative velocities, the supersonic stellar wind gets shocked and due to a non-zero relative speed, an elongated closed cavity forms separated by the shocked stellar-wind layer from the surrounding interstellar medium.

There are several indications that the Galactic centre environment is strongly magnetized at both larger and smaller scales \citep[see in particular][]{1987ApJ...320..545Y,2013Natur.501..391E,2018A&A...618L..10G}. Multiwavelength measurements of the Galactic centre magnetar involving the Faraday rotation measurement \citep{2013Natur.501..391E} revealed a strong magnetic field of $B\geq 50\,{\rm \mu G}$ at the projected distance scale of $\sim 3''=0.12\,{\rm pc}$. For the Galactic centre plasma, such a strong field is dynamically crucial and implies that the plasma at the Bondi radius is magnetically dominated. In combination with the number density of $n_{\rm e}^{B}=18.3 \pm 0.1\,{\rm cm^{-3}}$ at the Bondi radius \citep{2015A&A...581A..64R}, it is possible to estimate the Alfv\'en velocity at which disturbances propagate,
\begin{equation}
  v_{\rm A}=\sqrt{\frac{B^2}{\mu_0 \mu m_{\rm H}n_{\rm e}^{\rm out}}}\approx 36\left(\frac{B}{50\,{\mu G}} \right)\left(\frac{\mu_{\rm HII}}{0.5}\right)^{-1/2}\left(\frac{n_{\rm e}^{\rm out}}{18.3\,{\rm cm^{-3}}}\right)^{-1/2}\,{\rm km\,s^{-1}}\,.
  \label{eq_alfven_velocity}
\end{equation}
Dividing Eq.~\eqref{eq_orb_vel} by Eq.~\ref{eq_alfven_velocity} defines the Alfv\'enic Mach speed,
$\mathcal{M}_{\rm A}\equiv v_{\rm rel}/v_{\rm A}$, which turns out to be larger than unity in the magnetically dominated plasma. For a more detailed investigation, see also \citet{2017bhns.work..237Z} and Fig. 1 therein. 

Based on the theoretical estimates of the Mach speed, the formation of stellar bow shocks is expected in the innermost parsec of the Galactic centre. Several of them have been detected as comet-shaped structures in the NIR-, but more prominently in MIR-domains \citep{2002ApJ...575..860T,2003ANS...324..551R,2004ApJ...602..770G,2005ApJ...624..742T,2010A&A...521A..13M,2011A&A...534A.117B,2013A&A...557A..82B,2013A&A...551A..35R,2014A&A...567A..21S}, where the emission of dust of $\sim 1000\,{\rm K}$ becomes more prominent. Recently, we also showed how it is possible to constrain the bow-shock properties for fainter, unresolved NIR-excess sources in the inner S cluster \citep{2013A&A...551A..18E}, specifically for the fast-moving DSO/G2 source. This can be done by using the polarized signal in the NIR-domain in combination with the 3D Monte Carlo radiative transfer modelling \citep{2016A&A...593A.131S,2017A&A...602A.121Z}. The orientation of a bow shock and its characteristics (surface density, velocity field) also strongly depend on the presence and the magnitude of an inflow/outflow \citep{2016MNRAS.455.1257Z}. In particular, the presence of an ambient outflow of the order of $\sim 1000\,{\rm km\,s^{-1}}$ from the direction of Sgr~A*, which is inferred from the orientation and the size of comet-shaped sources X3 and X7 \citep{2010A&A...521A..13M}, induces an asymmetry in the evolution of the stellar bow-shock size and its other characteristics along the orbit of a stellar source.

 In this contribution, we analyze the potential of bow shocks in constraining the properties of the hot accretion flow around Sgr~A*. The article is structured as follows: in Section~\ref{overview} we provide an overview of already detected and studied bow-shock sources, subsequently in Section~\ref{probing_flow_stars} we summarize the basic aspects of constraining the accretion flow with stars, with the focus on using X-ray and radio flares (sub-Section~\ref{x-ray_radio_flares}) and on using bow-shock sizes at different locations along the stellar orbit to infer the density slope (sub-Section~\ref{density_slope}). This is extended in Section~\ref{statistics} for more bow-shock detections along the orbit of one star and for multiple bow-shock detections for several different sources. We provide a short summary in Section~\ref{summary}.  
 
\section{Overview of detected bow shock sources}
\label{overview}

In Table~\ref{tab_bow_shocks}, we list previously detected NIR- and MIR-sources that have been resolved and show an elongated, bow-shock morphology. In general, apart from the elongated morphology for detected sources, bow-shock sources exhibit a NIR-excess from NIR towards MIR wavelengths, which can be simply interpreted by the reprocessing of UV emission of a star by a surrounding dust \citep{2006ApJ...642..861V}. Bow shocks also exhibit a prominent linear polarization, whose degree rises from NIR to MIR wavelengths, which is caused by dust scattering, overall non-spherical geometry and a likely alignment of dust grains by magnetic field \citep{2013A&A...557A..82B}, which is a prominent feature in terms of linear polarization along the minispiral arms \citep{2018MNRAS.476..235R}.

\begin{table}[h!]
   \begin{tabular}{c|c|c|c}
     \hline
     \hline
     Name & Coordinates & Projected offset (Sgr~A*) & Publication\\
     \hline
     X3  & $17^{\rm h} 45^{\rm m} 39.85^{\rm s}$, $-29^{\circ} 00' 30.5''$ & $3.4''$ & [1] \\
     X7  & $17^{\rm h} 45^{\rm m} 40.00^{\rm s}$, $-29^{\circ} 00' 28.6''$ & $0.8''$ & [1] \\ 
     IRS 1W & $17^{\rm h} 45^{\rm m} 40.442^{\rm s}$, $-29^{\circ} 00' 27.51''$ & $5.3''$ & [2]\\
     IRS 5 & $17^{\rm h} 45^{\rm m} 40.700^{\rm s}$, $-29^{\circ} 00' 18.51''$ & $13.1''$ & [2]\\
     IRS 10W & $17^{\rm h} 45^{\rm m} 40.5305^{\rm s}$, $-29^{\circ} 00 23.001''$ & $8.3''$ & [2]\\
     IRS 21 & $17^{\rm h} 45^{\rm m} 40.222^{\rm s}$, $-29^{\circ} 00' 30.85''$ & $3.6''$ & [2]\\
     IRS 2L & $17^{\rm h} 45^{\rm m} 39.778^{\rm s}$, $-29^{\circ} 00' 32.11''$ & $5.2''$  & [3]\\
     IRS 8 & $17^{\rm h} 45^{\rm m} 40.14^{\rm s}$, $-28^{\circ} 59' 58.7''$ & $29.1''$ & [4], [5]\\ 
     X24 (MP-09-14.4)   & $17^{\rm h} 45^{\rm m} 39.424^{\rm s}$, $-29^{\circ} 00' 42.85''$ & $17''$ & [6], [7]\\ 
     IRS7 & $17^{\rm h} 45^{\rm m} 39.987^{\rm s}$, $-29^{\circ} 00' 22.24''$ & $6''-7''$ & [8]\\
     \hline
   \end{tabular}
   \caption{Summary of prominent bow-shocks sources in the inner parsec of the Galactic centre - columns summarize the designation, coordinates, projected offset from Sgr~A* ($1''$ corresponds to $\sim 0.04\,{\rm pc}$ at the Galactic centre), and the publication where the most recent data can be found: [1] \citet{2007A&A...469..993M}, [2] \citet{2014A&A...567A..21S}, [3] \citet{2006ApJ...642..861V}, [4] \citet{2004ApJ...602..770G}, [5] \citet{2013A&A...551A..35R}, [6] \citet{2014IAUS..303..150S}, [7] \citet{2009ApJ...699..186Z}, [8] \citet{1991ApJ...371L..59Y}.}
   \label{tab_bow_shocks}
\end{table} 

The importance of stellar bow shocks in the Galactic center lies in the fact that they can be employed as
\begin{itemize}
  \item[(a)] probes of circumnuclear ambient density and interstellar medium velocity in case the stellar parameters are known, which directly follows from the stand-off distance relation \citep{1996ApJ...459L..31W},
  \begin{equation}
  R_0 \sim \left(\frac{\dot{m}_{\rm w}v_{\rm w}}{4\pi \rho_{\rm a}v_{\rm rel}^{2}} \right)^{1/2}\,,
  \label{eq_standoff_distance0}
  \end{equation}
  where $\rho_{\rm a}$ is the ambient density, $v_{\rm rel}$ is the relative velocity of the star with respect to the ambient medium, $\dot{m}_{\rm w}$ is the mass-loss rate of the star and $v_{\rm w}$ is the terminal velocity of its stellar wind,
  \item[(b)] on the other hand, they can be used for estimating stellar properties in case the ambient medium density and its velocity are known alongside the stellar velocity, which again follows from Eq.~\eqref{eq_standoff_distance0},
  \item[(c)] combining approaches outlined in points (a) and (b), they can provide extra information on the dynamics of the nuclear star cluster. In particular, \citet{2014A&A...567A..21S} found that the bow-shock sources IRS 21, IRS 1W, IRS 5, and IRS 10W belong to a scattered population of young massive Wolf-Rayet stars, which are not a part of any coherent stellar structure previously known in the nuclear star cluster, in particular the clockwise stellar disc inside $0.5\,{\rm pc}$,
  \item[(d)] they can be used as probes of the hot accretion flow inside the Bondi radius of Sgr~A*, i.e. within the S cluster, as we will discuss in more detail below.   
\end{itemize}
 
\section{Probing accretion flow around Sgr~A* with stars}
\label{probing_flow_stars}

Compact radio source Sgr~A* is surrounded by one of the densest as well as the oldest stellar clusters in the Galaxy \citep{2014CQGra..31x4007S}. The stellar motion is dominated by the potential of Sgr~A* inside the sphere of the gravitational influence of the black hole, $r_{\rm h}\equiv GM_{\bullet}/\sigma_{\star}^2$, where $\sigma_{\star}$ is the one-dimensional stellar velocity dispersion, which can be distance-dependent. Therefore, kinematically it is easier to describe the SMBH potential dominance by the radius beyond which the mass of the central black hole is smaller than the enclosed stellar mass, $M_{\bullet}\lesssim M_{\rm enc}(r>r_{\rm h})$, which is about 2 pc for the Galactic centre Nuclear Stellar Cluster \citep{2013degn.book.....M}. Inside the influence radius, one can treat the stellar velocities as Keplerian as the first approximation, see Eq.~\eqref{eq_orb_vel}.

The hot ambient gas can be captured by Sgr~A* at smaller scales of $1''-2''$, which is given by the Bondi radius for steady, spherical accretion,

\begin{equation}
  R_{\rm B}=\frac{GM_{\bullet}}{c_{\rm s}^2}\,.
  \label{eq_bondi_radius}
\end{equation}
The accretion flow inside the Bondi radius and its associated radiative properties are generally given by radiatively inefficient accretion flow solutions \citep[RIAFs,][]{}. The RIAF region inside Bondi radius coincides with the innermost cluster of fast B-type stars -- so-called S cluster \citep{1996Natur.383..415E,1997MNRAS.284..576E} -- that is mostly located inside the innermost arcsecond, which corresponds to $\sim 0.04\,{\rm pc}$ for the Galactic centre. This provides an opportunity to test the accretion flow with stars by studying,
\begin{itemize}
  \item their potential bow-shocks as they interact with the surrounding medium, which could be revealed in the infrared, the X-ray or the radio domain, as was modelled by several authors \citep{2016MNRAS.459.2420C,2016MNRAS.455.1257Z,2017A&A...602A.121Z,2018MNRAS.479.5288B,2018A&A...616L...8S},
  \item longer X-ray flares which could result from stellar photons Compton-upscattered in the inner portions of the hot accretion flow \citep{2005A&A...429L..33N}.
\end{itemize}
In case the stellar bow shocks are resolved for the inner S stars, their properties are expected to vary quite significantly along their orbit. For instance, the stand-off distance given by Eq.~\eqref{eq_standoff_distance0} assuming the equilibrium between the ram and wind pressure, varies between the apocentre and the pericentre of the orbit as,
\begin{equation}
  \frac{R_{0A}}{R_{0P}}=\left(\frac{r_{\rm p}}{r_{\rm a}}\right)^{-\gamma/2} \left(\frac{v_{\rm \star p}}{v_{\rm \star a}}\right)\,,
  \label{eq_ratio_AP}
\end{equation} 
where $\gamma$ is the power-law slope of the accretion flow, whose number density is simply given as $n_{\rm a}=n_{0}(r/r_0)^{-\gamma}$ with $\gamma$>0, and in the further derivations, we assume that the stellar relative velocity can be approximated with the stellar velocity as given by the \textit{vis-viva} integral, $v_{\star}=\sqrt{GM_{\bullet}(2/r-1/a)}$. Then the ratio in Eq.~\eqref{eq_ratio_AP} can be expressed elegantly only as the function of the orbital eccentricity and the density slope, 

\begin{equation}
  \frac{R_{0A}}{R_{0P}}=\left(\frac{1+e}{1-e} \right)^{\gamma/2+1}\,.
  \label{eq_ratio_AP_e}  
\end{equation}

Similarly, the ratio between the \textit{latus-rectum} orbital position (the true anomaly of $90^{\circ}$) and the pericentre can simply be derived as,

\begin{equation}
   \frac{R_{0L}}{R_{0P}}=\frac{(1+e)^{\gamma/2+1}}{(1+e^2)^{1/2}}\,.
   \label{eq_ratio_LP_e}
\end{equation}

\begin{figure}[h!]
   \centering
   \includegraphics[width=\textwidth]{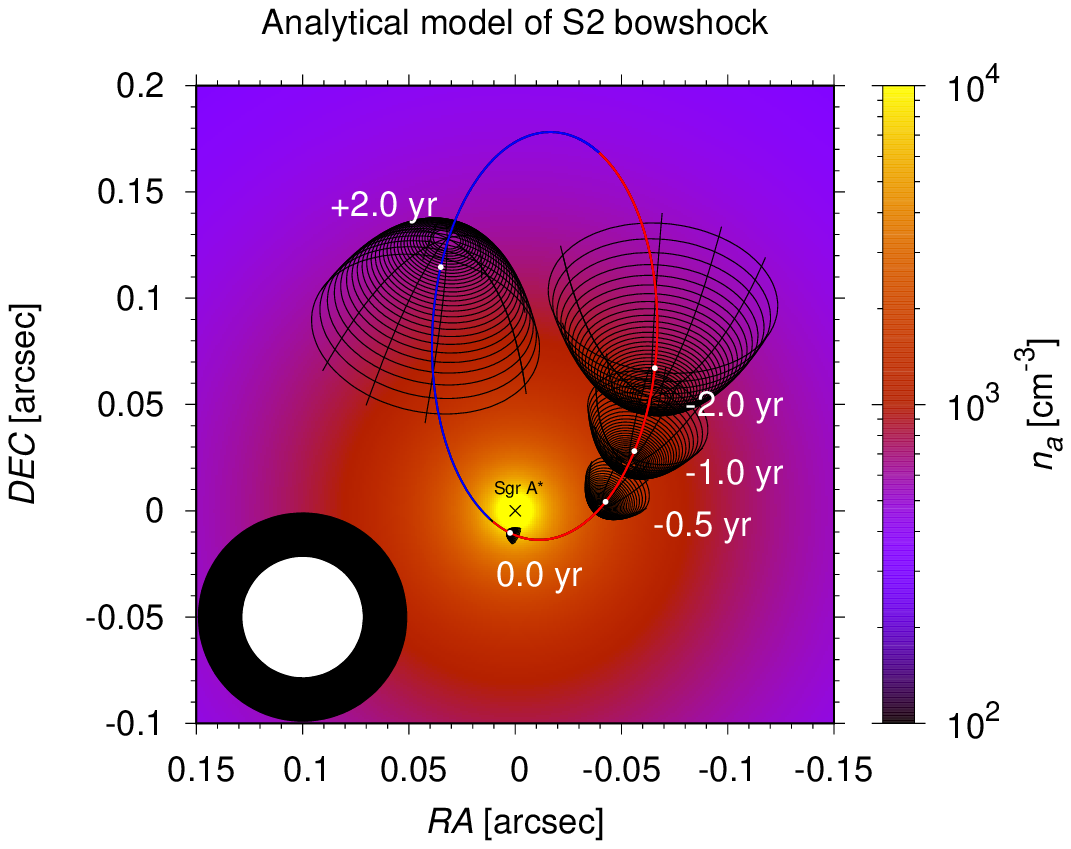}
    \caption{Bow-shock size evolution of S2 star orbiting Sgr~A* calculated according to the thin-shell analytical solution by \citet{1996ApJ...459L..31W}. Colour-coded background expresses RIAF density profile according to the solution by \citet{2003ApJ...598..301Y} and \citet{2006ApJ...640..319X}: $n_{\rm a}\approx 10^4(r/10^{15}\,{\rm cm})^{-1}\,{\rm cm^{-3}}$. Two concentric black and white circles in the lower left corner express the diffraction-limit resolution with eight-meter class telescopes in the NIR $L'$ and $K_{\rm s}$-band, respectively. For S2 star, the bow-shock would be unresolved close to the pericentre of its orbit, but it could be revealed via the spectral changes of Sgr~A*-S2 system during the pericentre passage.}
    \label{fig_S2_bowshock}
\end{figure}

Hence, it follows from Eqs.~\eqref{eq_ratio_AP_e} and \eqref{eq_ratio_LP_e} and for non-zero eccentricity, there is always a bow-shock size variation, even for a constant ambient density with $\gamma=0$. For instance, for the short-period S2 star whose eccentricity is currently well-constrained to be $e=0.884$, we get the apocentre--pericentre variation of $R_{0A}/R_{0P}\approx 65$ and the latus-rectum--pericenre variation of $R_{0L}/R_{0P}\approx 2$ for the inferred RIAF density slope close to $\gamma\approx 1$ \citep{2013Sci...341..981W}, which is also illustrated in Fig.~\ref{fig_S2_bowshock} for the analytical thin-shell bow-shock solution \citep[see][]{1996ApJ...459L..31W}. Since the radiative properties of bow shocks are proportional to their linear size, this motivates to search for potential flux density changes for high-eccentricity S stars, for which the changes along their orbit are most pronounced. According to Fig.~\ref{fig_S2_bowshock}, the bow-shock is the smallest at the pericentre of a stellar orbit due to the largest ram-pressure, but at the same time it will have the largest luminosity.   

\subsection{Inferring density from the (non)detection of radio and X-ray flares} 
\label{x-ray_radio_flares}

It was suggested using both analytical and numerical approaches that the interaction of the stellar wind of the fast-moving S2 star with the accretion flow should lead to the detectable month-long X-ray flare with the peak luminosity of $L_{\rm X}\approx 4\times 10^{33}\,{\rm erg\,s^{-1}}$ due to the increased thermal X-ray bremsstrahlung originating in the bow-shock \citep{2013MNRAS.433L..25G,2016MNRAS.459.2420C}. Such an increase is comparable to the quiescent X-ray emission of the Sgr~A* accretion flow and could be marginally detected with the current X-ray instruments. In principle, it could help constrain the accretion flow properties (density, density slope, its thickness) at the distance of the order of $1000$ Schwarzschild radii ($r_{\rm s}$), which is in general an unprobed, intermediate region \citep{2019ApJ...871..126G} between the Bondi radius region of $10^{5}\,r_{\rm s}$ probed in soft X-ray thermal emission \citep{2003ApJ...591..891B} and the region on the scale of $10-100\,r_{\rm s}$ probed in the submm-domain using the polarized emission \citep{2003ApJ...588..331B,2007ApJ...654L..57M}. Another contribution to the X-ray flux comes from the UV/optical photons of the S2 star that are Compton-upscattered in the inner hot RIAF region \citep{2005A&A...429L..33N}. Based on the non-detection of any month-long X-ray flare during S2 pericenter-passages in 2002 and 2018, the density structure and the slope of the accretion flow are fully consistent with the hot, diluted RIAF solution of \citet{2003ApJ...598..301Y} with the upper limit on the accretion rate of $\dot{M}_{\rm acc} \approx 2 \times 10^{-7}\,M_{\odot}{\rm yr^{-1}}$ \citep{2007ApJ...654L..57M}. This is also consistent with the updated 3D AMR hydrodynamical simulations of the interaction between the S2 stellar wind and the accretion flow, which predicted no significant X-ray emission excess close to the S2 pericentre passage \citep{2018A&A...616L...8S}.

The detected dusty source G2 or DSO \citep{2012Natur.481...51G,2013A&A...551A..18E} was also predicted to drive a shock into the ambient medium, which in principle could lead to the detectable nonthermal radio synchrotron emission \citep{2012ApJ...757L..20N}. This theory can be expanded in a straightforward way to all S stars, including S2 \citep{2016MNRAS.455L..21G}. The general prediction was that the electrons in the accretion flow are accelerated in the bow-shock region to relativistic energies, with the peak flux close to $1\,{\rm GHz}$. Concerning stellar bow shocks, the main sources of uncertainty are the bow-shock size and the magnetic field strength that is enhanced in the shocked region. However, for realistic values of the parameters applied to DSO/G2 case, the radio synchrotron emission was found to be well below the quiescent radio emission of Sgr~A* \citep{2013MNRAS.436.1955C,2016MNRAS.455.1257Z}. For S2 star, \citet{2016MNRAS.455L..21G} show that its synchrotron flux would be comparable to that of Sgr~A* at 10 GHz (radio band) as well as $10^{14}\,{\rm Hz}$ (infrared) for extreme combinations of the mass-loss rate and the wind velocity: $(\dot{m}_{\rm w}, v_{\rm w})=(10^{-5}\,M_{\odot}{\rm yr^{-1}},1000\,{\rm km\,s^{-1}})$ and $(\dot{m}_{\rm w}, v_{\rm w})=(10^{-6}\,M_{\odot}{\rm yr^{-1}},4000\,{\rm km\,s^{-1}})$.  

\subsection{Density slope determination for one bow-shock source orbiting Sgr~A*}
\label{density_slope}

The bow-shock length-scale is given by the stagnation (stand-off) distance $R_0$, which can be derived under the assumption that in the stationary case, the stellar wind pressure is equal to the sum of the ram pressure and the thermal pressure, $P_{\rm w}=P_{\rm ram}+P_{\rm th}=P_{\rm ram}(1+\alpha)$, where the last equality is obtained using the ratio $\alpha=P_{\rm th}/P_{\rm ram}$. In the most general case, the stagnation distance of the bow-shock, i.e. the distance from the star to the tip of the bow shock, can be expressed as \citep{1996ApJ...459L..31W,1997ApJ...474..719Z,2016MNRAS.459.2420C},

\begin{equation}
  R_0=\left(\frac{\dot{m}_{\rm w}v_{\rm w}}{\Omega_{\rm w}\rho_{\rm a} v_{\rm rel}^2(1+\alpha)}\right)^{1/2}\simeq C_{\star}\rho_{\rm a}^{-1/2} v_{\rm rel}^{-1}\,,
  \label{eq_standoff}
\end{equation} 
where the last equality is obtained under the assumption that the thermal pressure is negligible, i.e. $\alpha \rightarrow 0$, which is essentially met for the case of highly-supersonic motion, 
$\mathcal{M}\equiv v_{\rm rel}/c_{\rm s}=1/\sqrt{\nu\alpha} \gg 1$ for $\alpha \rightarrow 0$ ($\nu$ is the adiabatic index). The quantity $C_{\star}^2=\dot{m}_{\rm w}v_{\rm w}/\Omega_{\rm w}$ represents the momentum flux of the stellar wind per unit solid angle $\Omega_{\rm w}$, which is equal to $\Omega_{\rm w}=4\pi$ for an isotropic outflow. The parameter $C_{\star}$ may be assumed to be constant for several stellar orbits. We estimate the typical value for the brightest star in the S cluster - S2 based on its stellar parameters \citep{2008ApJ...672L.119M},

\begin{equation}
  C_{\star}^2=7.85\times 10^{-5}\left(\frac{\dot{m}_{\rm w}}{10^{-7}\,M_{\odot}\,{\rm yr^{-1}}}\right)\left(\frac{v_{\rm w}}{1000\,{\rm km\,s^{-1}}}\right)\, M_{\odot}\,{\rm km\,s^{-1}\,yr^{-1}\,sr^{-1}}\,.
  \label{eq_momentum_flux_S2}
\end{equation}

In the further analysis, we neglect the thermal pressure as well as the motion of the ambient medium. Hence, the relative velocity can be approximated by stellar orbital velocity, $v_{\rm rel} \simeq  v_{\star}$. The ratio of the bow-shock stand-off distances, $R_{01}$ and $R_{02}$, for two corresponding distances of the star from Sgr~A*, $r_1$ and $r_2$ (where $r_1 > r_2$), can then be expressed as, 

\begin{equation}
  \frac{R_{01}(r_1)}{R_{02}(r_2)}=\left(\frac{n_{\rm a2}}{n_{\rm a1}}\right)^{1/2}\left(\frac{v_{\rm rel2}}{v_{\rm rel1}}\right)=\left(\frac{r_{1}}{r_{2}}\right)^{\gamma/2}\left(\frac{v_{\star 2}}{v_{\star 1}}\right)\,,
  \label{eq_ratio1}
\end{equation} 
where we adopted the power-law density profile for the ambient medium, $\rho_{\rm a}=\mu m_{\rm H} n_0(r/r_0)^{-\gamma}$, where $\gamma>0$ is the power-law slope. 

The ratio expressed by Eq.~\eqref{eq_ratio1} can be further expanded using the general relation for the Keplerian orbital velocity along the elliptical orbit, $v_{\star}=\sqrt{GM_{\bullet}(2/r-1/a)}$, where $r$ is the radial distance from Sgr~A* and $a$ is a semi-major axis of the elliptical orbit. We get,

\begin{equation}
  \frac{R_{01}}{R_{02}}(r_1,r_2,a,\gamma)=\left(\frac{r_{1}}{r_{2}}\right)^{(\gamma+1)/2}\left(\frac{2a-r_2}{2a-r_1}\right)^{1/2}\,.
  \label{eq_ratio2}
\end{equation} 

It is then trivial to express the power-law slope from Eq.~\eqref{eq_ratio2}. We denote it as $\gamma_{12}$ as it is essentially based on bow-shock sizes at two points,
\begin{equation}
 \gamma_{12}=\frac{2\log{(R_{01}/R_{02})}+\log{[(2a-r_1)/(2a-r_2)]}}{\log(r_1/r_2)}-1\,.
 \label{eq_gamma12}
\end{equation}

\section{Statistical approach}
\label{statistics}

Now we will present an analytical method for determining the density power-law slope of the ambient medium, in our case hot gas close to the Galactic centre, for multiple bow-shock detections. There are two plausible observable cases:
\begin{itemize}
 \item[(i)] multiple bow-shock detections for one star (multiple epochs along one orbit),
 \item[(ii)] multiple bow shock detections for different stars (single epoch sufficient). 
\end{itemize}

\begin{figure*}[tbh]
 \centering
 \begin{tabular}{cc}
 \includegraphics[width=0.5\textwidth]{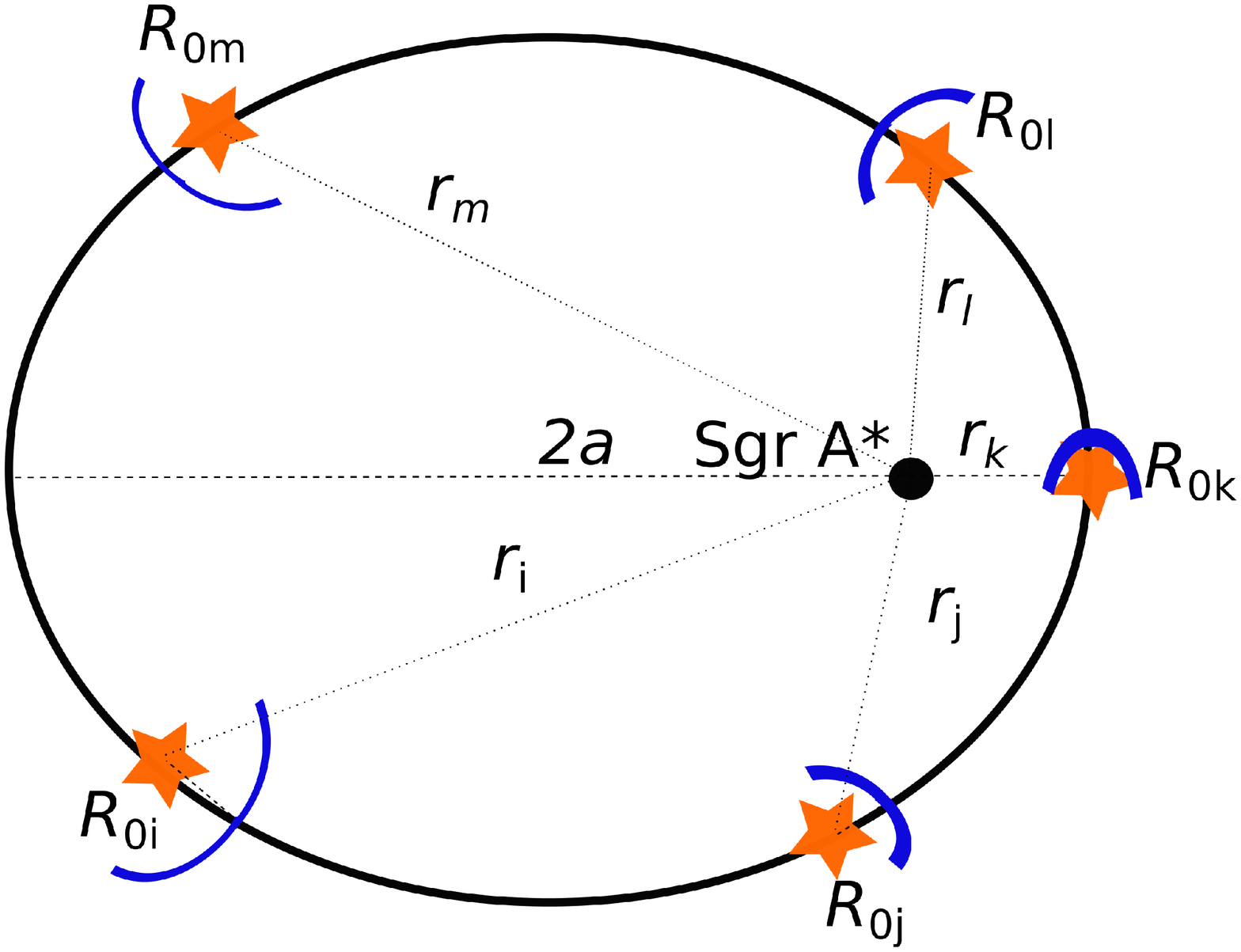} &
 \includegraphics[width=0.5\textwidth]{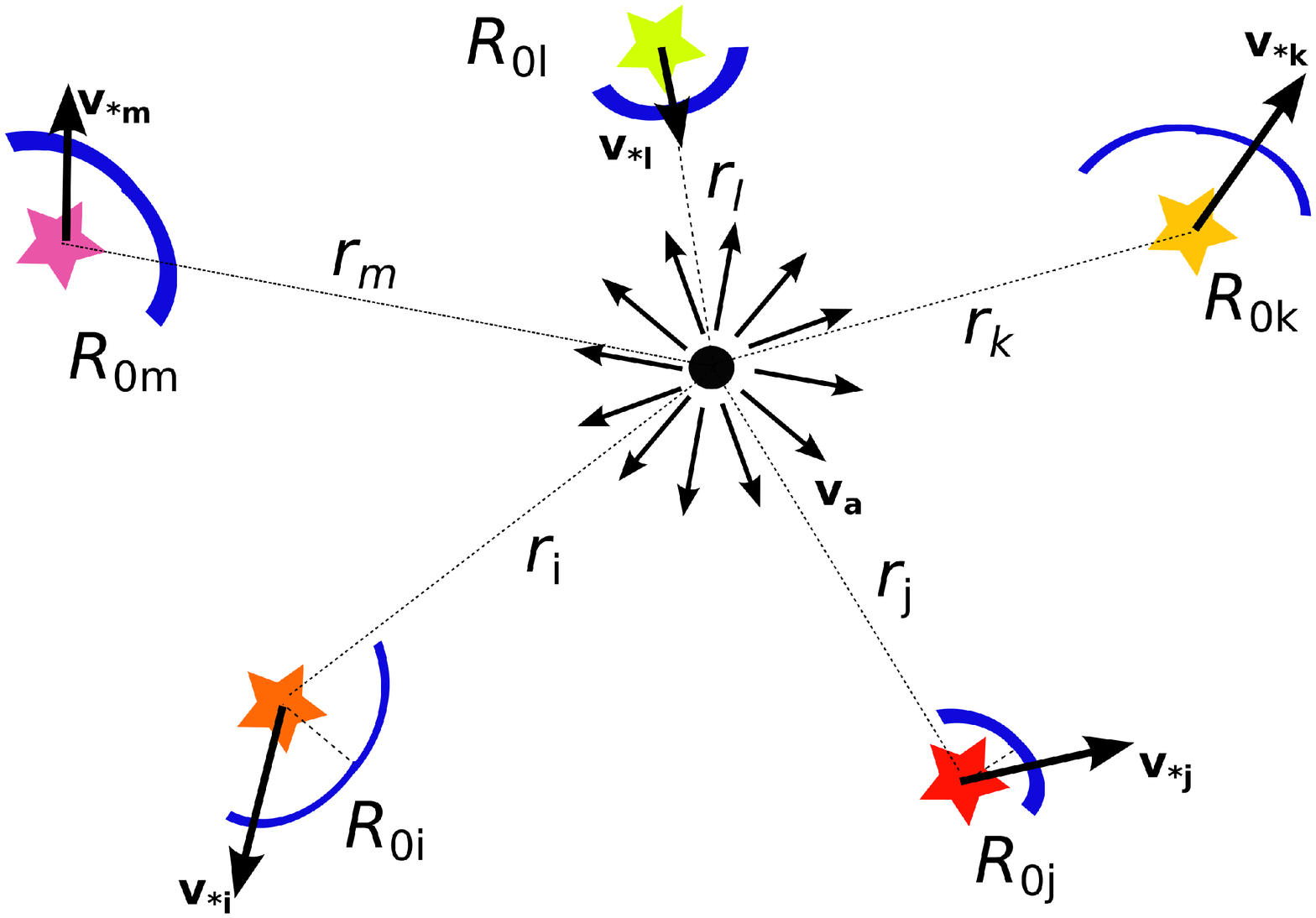}
 \end{tabular}
 \caption{Illustration of a statistical approach for constraining the power-law slope of a spherical density distribution close to Sgr~A*: \textit{Left panel:} Multiple bow shock detections of one star along an elliptical orbit. \textit{Right panel:} Multiple bow shock detections for different stars with different mass-loss rates $\dot{m}_{{\rm w}i}$ and terminal stellar wind velocities $v_{{\rm w}i}$. In this case also deprojected stellar orbital velocities $\mathbf{v_{\star i}}$ and the outflow vector field $\mathbf{v_{{\rm a}}}$ are labelled for illustration.}
 \label{fig_statistics}
\end{figure*}

In Fig~\ref{fig_statistics} we illustrate both cases -- (i) one star (left panel) and (ii) more bow shocks associated with different stars (right panel).  

This will make the analysis more practical in terms of actual observations. In the following analysis, we will also include a potential non-negligible motion of the ambient medium described in general by a vector field $\mathbf{v}_{\rm{a}}(r,\theta,\phi)$. Hence, the relative velocity of the star with respect to the medium is $\mathbf{v_{\rm rel}}=\mathbf{v_{\star}}-\mathbf{v_{\rm a}}$ and the magnitude is $v_{\rm rel}=\sqrt{\mathbf{v_{\rm rel}} \cdot \mathbf{v_{\rm rel}}}=\sqrt{v_{\star}^2+v_{\rm a}^2-2v_{\star}v_{\rm a}\cos{\zeta}}$, where $\zeta$ is the angle between the velocity vector of the star and the medium at a given point.    

\subsection{Multiple bow-shock detections for one star}

Let us suppose that the bow shock was detected and its deprojected standoff distance $R_{0i}$ inferred for $N$ locations along the orbit. Then we can calculate the density slope $\gamma_{ij}$ for any two locations $i$ and $j$, where $\{i,j\}\in \{1,\ldots,N\}$, according to Eq~\eqref{eq_gamma12},

\begin{equation}
 \gamma_{ij}=\frac{2 \log{[R_{0i}(r_i)/R_{0j}(r_j)]}+\log{[(2a-r_i)/(2a-r_j)]}}{\log{(r_i/r_j)}}-1\,. 
 \label{eq_power_law_index_ij_onestar}
\end{equation}

When the ambient medium has a known non-negligible velocity field $\mathbf{v_{\rm{a}}}$ with respect to the star, the formula~\eqref{eq_power_law_index_ij_onestar} needs to be modified into a more general form,

\begin{equation}
 \gamma_{ij}=\frac{2 \log{[R_{0i}(r_i)/R_{0j}(r_j)]}+\log{\left[\frac{v_{\star i}^2+v_{{\rm a }i}^2-2v_{\star i}v_{{\rm a }i}\cos{\zeta_i}}{v_{\star j}^2+v_{{\rm a }j}^2-2v_{\star j}v_{{\rm a }j}\cos{\zeta_j}}\right]}}{\log{(r_i/r_j)}}-1\,,
 \label{eq_power_law_index_ij_onestar_general}
\end{equation}
where $v_{\star i}$, $v_{{\rm a}i}$, and $\zeta_{i}$ correspond to the stellar orbital velocity, ambient velocity, and the angle between the two velocity vectors, respectively, at position $i$.

For $N$ bow-shock measurements, $N\geq 2$, we have $N(N-1)/2$ indices $\gamma_{ij}$, from which we can calculate the mean value,

\begin{equation}
 \overline{\gamma}=\frac{2}{N(N-1)}  \sum_{\substack{i,j=1\\
                   j\neq i}}^N \gamma_{ij}\,,
 \label{eq_gamma_mean_one_star}                  
\end{equation}

\subsection{Multiple bow-shock detections for different stars}

Since due to large stellar densities in the innermost arcsecond it is difficult to detect a bow shock, a currently more accessible method is to use more distant bow-shock sources at different locations at a single epoch. However, a disadvantage is that for $N$ bow shock sources with approximately isotropic stellar winds, we have $2N$ more parameters $(\dot{m}_{\rm{w}},v_{\rm{w}})$ than for a single source.

Having $i=1,\ldots,N$ detected bow shocks at deprojected distances $r_{i}$ from the Galactic centre, for the ratio of standoff radii $R_{\star i}$ and $R_{\star j}$ for any two sources the following relation holds,

\begin{equation}
  \frac{R_{\star i}}{R_{\star j}}=\frac{C_{\star i}}{C_{\star j}} \left(\frac{n_{\rm a}(r_j)}{n_{\rm a}(r_i)} \right)^{1/2} \left(\frac{v_{{\rm rel}j}}{v_{{\rm rel}i}} \right)\,.  
\end{equation}

As in previous cases, the power-law index $\gamma_{ij}$ may be determined as follows,
\begin{equation}
 \gamma_{ij}=\frac{2 [\log{(R_{\star i}/R_{\star j})}-\log{(C_{\star i}/C_{\star j})}]+\log{\left[\frac{v_{\star i}^2+v_{{\rm a }i}^2-2v_{\star i}v_{{\rm a }i}\cos{\zeta_i}}{v_{\star j}^2+v_{{\rm a }j}^2-2v_{\star j}v_{{\rm a }j}\cos{\zeta_j}}\right]}}{\log{(r_i/r_j)}}\,.
 \label{eq_power_law_index_ij_multiple}
\end{equation}

In case the ambient velocity field is negligible in comparison with stellar velocities, $\mathbf{v_{{\rm a}}}=0\,{\rm km\,s^{-1}}$, and the orbital velocities at $r_{i}$ are approximately given by Keplerian circular velocities, $v_i=(GM_{\bullet}/r_i)^{1/2}$, relation~\eqref{eq_power_law_index_ij_multiple} may be simplified into the form, 

\begin{equation}
 \gamma_{ij}=\frac{2 [\log{(R_{\star i}/R_{\star j})}-\log{(C_{\star i}/C_{\star j})}]}{\log{(r_i/r_j)}}-1\,.
 \label{eq_power_law_index_ij_multiple_simplified}
\end{equation}

The mean value of the density power-law profile $\overline{\gamma}$ for $N$ different bow shocks is determined by the relation that is identical to Eq.~\eqref{eq_gamma_mean_one_star}. 

\section{Summary}
\label{summary}

In this contribution, we presented an overview of observational studies of bow-shock sources in the Galactic centre region. About ten stellar bow shocks have been detected and analyzed in the NIR- and MIR-domain where they are the most prominent because of thermal dust emission and scattering. In addition, we outlined that their signatures should also be detected in the X-ray domain due to thermal bremsstrahlung as well as in the radio domain due to non-thermal synchrotron emission. However, because of diluted ambient medium in the central parsec of our Galaxy, there have been no successful detections in these domains.

 Furthermore, we introduced a simple statistical method to determine the ambient density gradient in the inner stellar cluster region based on either multiple bow-shock detections for one star bound to Sgr~A* or bow-shock detections for more stars at different distances from Sgr~A*. Although the method is still observationally challenging, since several parameters need to be obtained (at least stand-off radii of bow shocks at different positions and corresponding deprojected radii), stellar bow shocks may become complementary probes of the hot accretion flow around Sgr~A* in the era of high-angular resolution when $30$-meter telescopes in the NIR- and MIR-domain are operational.  
 
 \section*{Acknowledgements}

 Michal Zaja\v{c}ek acknowledges the financial support from the National Science Centre, Poland, grant No. 2017/26/A/ST9/00756 (Maestro 9).

\bibliographystyle{aa} 
\bibliography{zajacek} 


\end{document}